\documentclass[aps,prl,twocolumn,showpacs,preprintnumbers,amsmath,amssymb,superscriptaddress]{revtex4-1}
\usepackage{graphicx}% Include figure files
\usepackage{dcolumn}% Align table columns on decimal point
\usepackage{bm}% bold math
\newcommand{\nc}{\newcommand}
\nc{\be}{\begin{equation}}
\nc{\ee}{\end{equation}}
\nc{\bea}{\begin{eqnarray}}
\nc{\eea}{\end{eqnarray}}
\nc{\bean}{\begin{eqnarray*}}
\nc{\eean}{\end{eqnarray*}}
\nc{\mb}{\mbox}
\nc{\rnc}{\renewcommand}
\nc{\vk}{\mb{\bf k}}
\nc{\vp}{\mb{\bf p}}
\nc{\vn}{\mb{\bf n}}
\nc{\vq}{\mb{\bf q}}
\nc{\rr}{\mb{\bf r}}
\nc{\vz}{\hat {\mb{\bf z}}}
\nc{\vj}{\mb{\boldmath$j$}}
\nc{\vg}{\mb{\boldmath$g$}}
\nc{\x}{\mb{\boldmath$x$}}
\nc{\A}{\mb{\boldmath$A$}}
\nc{\va}{\mb{\boldmath$a$}}
\nc{\vs}{\mb{\boldmath$\sigma$}}
\nc{\vpi}{\mb{\boldmath$\pi$}}
\nc{\nab}{\nabla}
\nc{\X}{\sf x}

\begin{document}

\title{Counter-propagating Fractional Hall states in mirror-symmetric Dirac semi-metals}

\author{Yafis Barlas}
\affiliation{Department of Electrical and Computer Engineering, University of California,
Riverside, California}
\thanks{Current Address: Department of Physics, Yeshiva University, New York, NY 10016, USA}
%\affiliation{Department of Physics and Astronomy, University of California,
%Riverside, California 92521, USA}

%--Outline
% 1) Introduction.
% 2) Quantum mirror Hall effects in Dirac semi-metals.
% 3) Fractional Hall effects protected by mirror symmetry (Laughlin wavefunction and hiearchy).  
% 4) Chern Simons Landau Ginzburg theory.
% 5) Edge state transition to a fractional mirror insulator.
% 6) Conclusion and Outlook

\begin{abstract}
The Landau bands of mirror symmetric 2D Dirac semi-metals (for example odd-layers of ABA-graphene) can be identified by their parity with respect to mirror symmetry. This symmetry facilitates a new class of counter-propagating Hall states. We predict the presence of a Laughlin-like correlated liquid state, at the charge neutrality point, with opposite but equal electron and hole filling factors $|\nu_{\pm}|=1/m$ ({\it m} odd). This state exhibits fractionally charged quasi-particle/hole pair excitations and counter-propagating edge states with opposite parity. Using a bosonized one-dimensional edge state theory, we show that the fractionally quantized two-terminal longitudinal conductance, $\sigma_{xx} = 2e^2/(m h)$, is robust to short-ranged inter-mode interactions.
\end{abstract}


\maketitle

The quantum spin Hall (QSH) state~\cite{PhysRevLett.95.226801,PhysRevLett.96.106802}, is characterized by one-dimensional helical edge states. This results in a quantized longitudinal two-terminal resistance~\cite{Konig766} along with a vanishing Hall resistance. Time reversal symmetry in the QSH state is essential, as it forbids back-scattering of the helical edge modes. Such symmetry protected topological (SPT) states can also be realized in Dirac semi-metals (graphene and bilayer graphene) at neutral charge density in the quantum Hall (QH) regime. Interactions in the $\nu =0$ QH state in graphene and bilayer graphene result in an insulating canted anti-ferromagnetic (CAF) state~\cite{PhysRevB.85.155439}, due to spontaneous ordering of the half-filled spin and valley degenerate Landau level. As the in-plane magnetic field is increased, this insulating CAF state transitions to a ferromagnetic state with counter-propagating spin polarized edge modes and a quantized two-terminal conductance~\cite{PhysRevB.85.155439,Young2014,Maher2013}. Here, the robust quantization requires spin rotational symmetry. Other interaction induced symmetry protected topological (SPT) states, like fractional topological insulators~\cite{PhysRevLett.103.196803} have also been proposed, but not realized experimentally. In this Letter, we propose a new class of interacting and non-interacting SPT phases protected by mirror symmetry in 2D Dirac semi-metals. 
%This state exhibits zero Hall conductance, but a quantized two-terminal longitudinal conductance at neutral charge density. 

In 2D Dirac semi-metals, Fermi-surface (FS) topology at neutral charge density can lead to two distinct types of semi-metallic band structures. Since the conduction band minima and valance band maxima must coincide in momentum space, this gives only two distinct possibilities for FS topology. At the charge neutrality point (CNP), either i) the FS consists of singular points in the Brillouin zone (BZ) (for example graphene and bilayer graphene bands), or ii) non-zero electron and hole pockets with, $n_{e} = -n_{h}$, co-exist at neutral charge density. The latter FS topology is only robust to perturbations, if the spinless electron and hole pockets belong to different irreducible representations of the lattice symmetry. The second scenario can be viewed as the analogue of band inversion in Dirac semi-metals. It is satisfied in the ABA-stacked trilayer graphene~\cite{PhysRevX.2.011004,PhysRevB.81.115315}. 

\begin{figure}
\begin{tabular}{c}
\includegraphics[clip,width=1.0\linewidth]{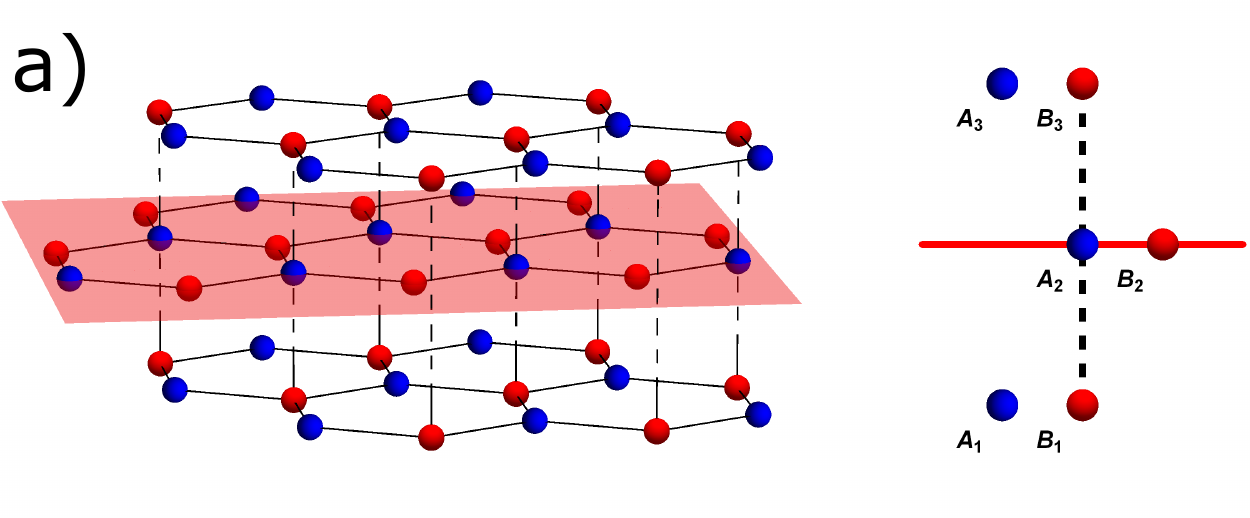}\\
\includegraphics[clip,width=1.0\linewidth]{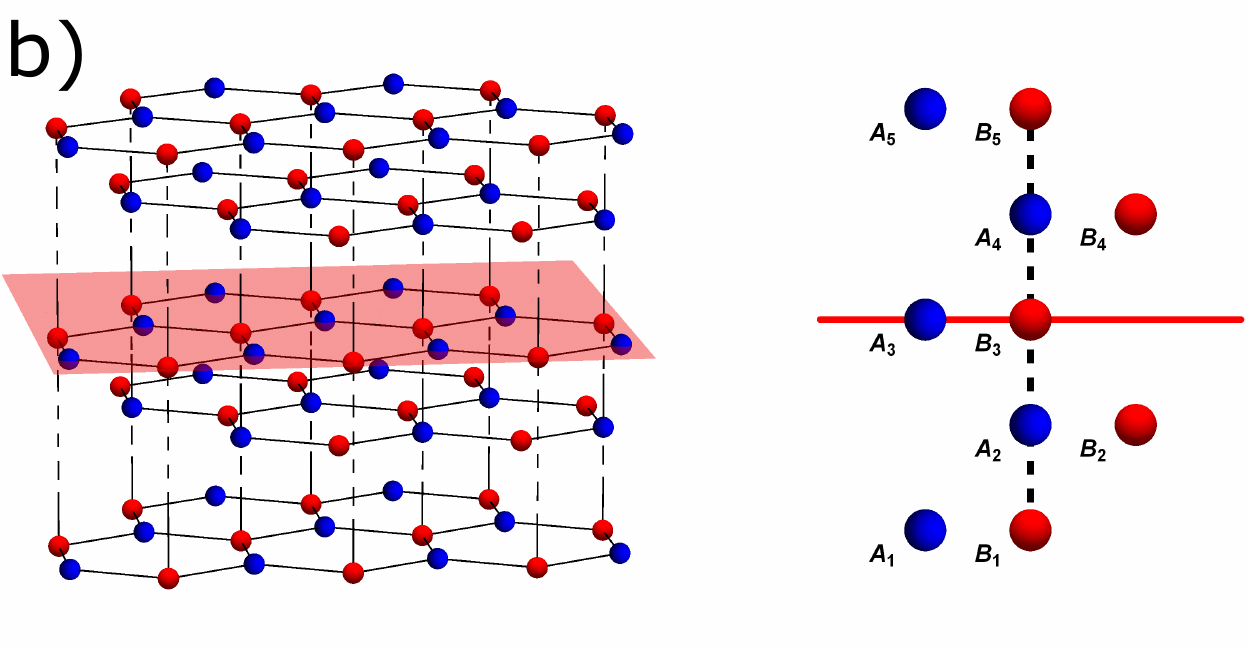}
\end{tabular}
\caption{Lattice structure of ABA-trilayer graphene (TLG) (a) and ABA-pentalayer graphene (b) with mirror symmetry about the middle layer  denoted by the red line (plane). Under mirror symmetry $A_{1} \leftrightarrow A_{3}$, $B_{1} \leftrightarrow B_{3}$, whereas the middle layer remains invariant for ABA-TLG.} 
\label{figureone}
\end{figure}

Layered graphene stacks, held together by van der Waals forces, add an extra degree of richness to graphene's electronic properties~\cite{PhysRevLett.97.036803,PhysRevB.73.245426,PhysRevB.77.155416}. One such stacking configuration is the ABA-stacked multilayer graphene. In this configuration, each layer has the same in-plane projection as its next nearest neighboring layer, implying that all next nearest layers are exactly aligned but vertically displaced. Additionally, each nearest neighboring layer is Bernal stacked, such that half of the atoms in alternate layers lie directly over the center of the hexagon while the other half lie directly over the atom in the nearest neighboring layer (see Fig.~\ref{figureone} (a) \& (b)). It follows that the even and odd layer stacks belong to different symmetry groups. Even N-layer ABA-multilayer graphene stacks are inversion symmetric (i.e. $\vec{r} \to -\vec{r}$), whereas odd N-layer ABA-multilayer graphene stacks satisfy mirror reflection symmetry (i.e. $(x,y,z) \to (x,y,-z)$). Therefore, the energy bands of odd N-layer graphene sheets can be classified in terms of their parity $(\eta = \pm)$ with respect to mirror reflection~\cite{PhysRevB.81.115315}. Furthermore, because of this lattice symmetry, the Hamiltonian of odd N-layer ABA-multilayer graphene becomes block diagonal in terms of the parity eigenstates.   
% Mirror symmetry is valid in the absence of an external potential difference $E_{\perp}$.

To make the discussion more precise, consider the simplest mirror symmetric 2D Dirac semi-metal: ABA-trilayer graphene (TLG). ABA-TLG is invariant under the $D_{3h}$ point group, which includes mirror symmetry about the middle layer. From now on we denote the inequivalent atomic sites of the $i^{th}$ graphene layer by $A_{i}$ and $B_{i}$. The lattice stacking is such that only half of the sub-lattice sites in each layer ($B_{1}, A_{2}, B_{3}$) have a near-neighbor in the adjacent layer, whereas the other half ($A_{1}, B_{2}, A_{3}$) don't have a near neighbor in the adjacent layers (see Fig.~\ref{figureone} a). Under mirror symmetry the sub-lattices on the top layer are interchanged with the sub-lattices in the bottom layer, $\psi_{A_{1}} \leftrightarrow \psi_{A_{3}}$ and $\psi_{B_{1}} \leftrightarrow \psi_{B_{3}}$, whereas the sub-lattices of the middle layer remain unchanged, $\psi_{A_{2}} \leftrightarrow \psi_{A_{2}}$ and $\psi_{B_{2}} \leftrightarrow \psi_{B_{2}}$.

In general, the mirror symmetry operator $\mathcal{M}$ can be expressed as a rotation by $\pi$ with the axis of rotation perpendicular to the mirror plane, followed by an inversion. This gives $\mathcal{M}= P D(\pi)$, where $P$ is the inversion operator which sends $\vec{r} \to - \vec{r}$, and $D(\pi)$ is the rotation operator which acts on the internal degrees of freedom such as spin. For spinless particles, this rotation matrix acts trivially and is equal to the identity matrix. The mirror symmetry operator satisfies $\mathcal{M}^2=1$ with eigenvalues $\pm 1$, corresponding to the even and odd parity. For ABA-TLG the sub-lattice orbital combinations $A_{\pm} = (A_{1} \pm A_{3})/\sqrt{2} $ and $B_{\pm} = (B_{1} \pm B_{3})/\sqrt{2}$, form the irreducible representations of this mirror symmetry. 
$A_{\pm}$ and $B_{\pm}$ have $+(-)$ even (odd) parity with respect to this mirror symmetry, while the middle layer $A_{2}$ and $B_2$ orbitals have even parity~\cite{PhysRevLett.97.036803,PhysRevB.87.115422}.

Therefore, the Hamiltonian for ABA-TLG can be separated into contributions from even and odd parity Hamiltonians, $\mathcal{H} = \mathcal{H}_{-} \otimes \mathcal{H}_{+}$~\cite{PhysRevB.87.115422}. The energy bands of the odd parity orbitals $(A_{-},B_{-}) $ belong to the class of gapped Dirac-like dispersion with,  
\begin{equation}
\label{Eq:oddHam}
\mathcal{H}_{-} = \hbar v (\hat{\sigma}_{x} \pi_{x} + \tau_{z} \hat{\sigma}_{y} \pi_{y}) + m_{-} \hat{\sigma}_{z} + \frac{\Delta}{2} \hat{\sigma}_{0},
\end{equation}
where $\sigma_{i}$'s denote the Pauli matrices and $\sigma_0 = \mathbb{I}_{2}$. Eq. (\ref{Eq:oddHam}) is acts on the odd-parity spinors, $\psi^{\dagger}_{o,\tau} = (\phi^{\dagger}_{A_{-},\tau},\phi^{\dagger}_{B_{-},\tau})$, $\tau = \pm $ denotes the $\bf {K} (\bf{K}')$ valley in the hexagonal BZ, $\hbar v = 688 $ meV nm, 
$\pi_{i} = i \partial_{i} - e A_{i} $ is the momentum operator in a high magnetic field $ B\hat{z} = \nabla \times A$, satisfying $[\pi_{x},\pi_{y}] = -i/l_{B}^2$, where $l_{B} = \sqrt{\hbar/eB}$ is the magnetic length. The mass gaps $\Delta=27 $ meV  and $m_{-}=3 $ meV are determined by the remote inter-layer hopping parameters and energy differences between stacked and non-stacked atoms~\cite{PhysRevLett.117.076807}. 

\begin{figure}[t]
%\begin{tabular}{cc}
%\includegraphics[clip,width=0.9\linewidth]{Fig1crop.pdf}\\
\includegraphics[width=1.0\linewidth]{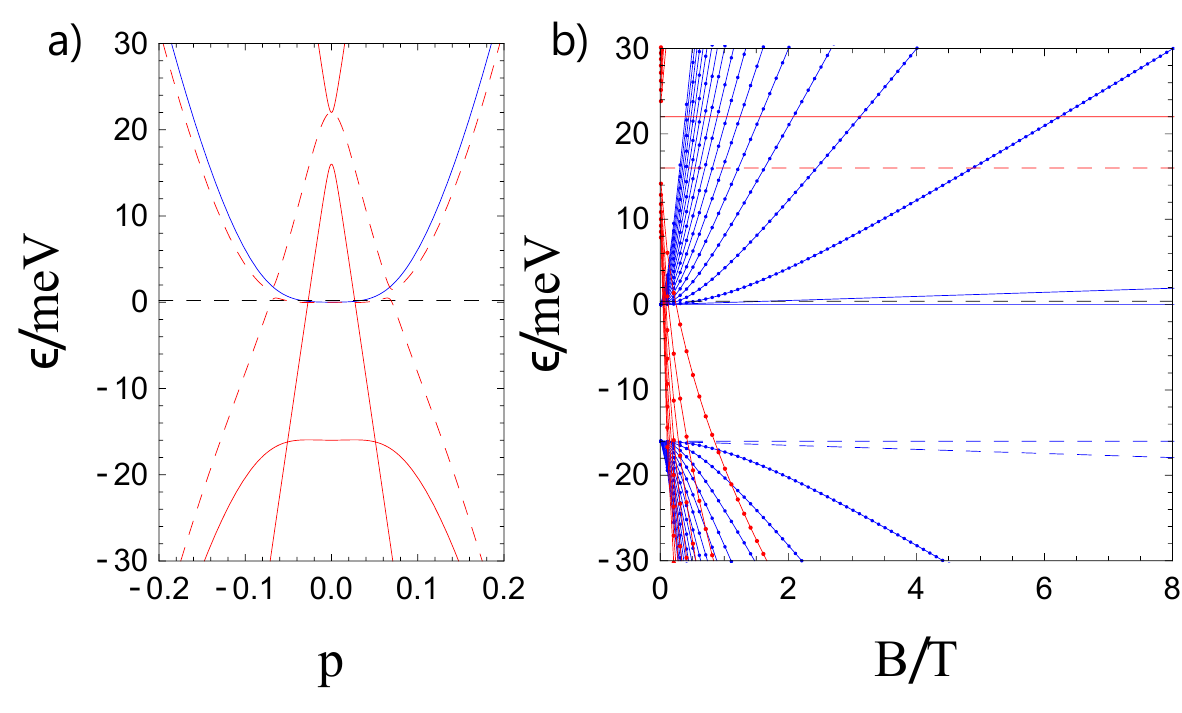}
%\end{tabular}
\caption{a) Energy dispersion of ABA-TLG at $E_{\perp} =0$ (solid line) and $E_{\perp}=40 {\rm meV}$ (dashed line). The semi-metallic phase requires $|\Delta| > (m_{+}+m_{-})$. (b) Landau level spectrum of ABA-TLG. In both figures red represents the odd parity electrons and blue denotes the even parity electrons and the black dotted line denotes the Fermi energy at the CNP. The dotted and solid lines denote different valleys (sub-lattices for the zLLs). At neutral charge density ($\nu = 0$) the even parity filling factor is $\nu_{+} =2$, while the odd parity filling factor is $\nu_{-} =-2$.} 
\label{figuretwo}
\end{figure}

The even parity orbitals $(A_{+},B_{2},A_{2},B_{+})$, exhibit a band dispersion similar to gapped bilayer graphene. At low-energies the orbitals $A_{2}$ and $B_{+}$ are pushed to high energies due to the direct inter-layer hopping $\gamma_{1} \sim 0.31$ eV. The resulting low-energy even parity Hamiltonian~\footnote{To keep the discussion simple, we express $\mathcal{H}_{+}$ within second order perturbation theory.}, $\mathcal{H}_{+}$ % = \mathcal{H}^{(0)}_{+} $% + \mathcal{H}^{(1)}_{+}$, 
acts on the even-parity spinors $\psi^{\dagger}_{e,\tau} = (\phi^{\dagger}_{A_{+},\tau},\phi^{\dagger}_{B_{2},\tau})$~\cite{PhysRevB.87.115422}, 
\begin{equation}
\mathcal{H}_{+} =  % \frac{v^2}{\sqrt{2} \gamma_{1}} 
\frac{-\hbar^2 v^2}{\sqrt{2} \gamma_{1}}
\bigg( \hat{\sigma}_{x} (\pi^2_{x} - \pi^2_{y}) + \tau_{z} \hat{\sigma}_{y} \{\pi_{x},\pi_{y}\} \bigg) + m_{+} \hat{\sigma}_{z}
- \frac{\Delta}{2}\hat{\sigma}_{0},
\end{equation}
here $ \{...\} $ denotes the anti-commutator with 
%$m = \hbar^2 v^2/\sqrt{2} \gamma_{1}$ is the effective mass for the even parity bands, 
$m_{+}=8$ meV~\cite{PhysRevLett.117.076807}. To ensure the semi-metallic FS topology (see in Fig~\ref{figuretwo} a), electron-hole band overlap at the CNP requires $|\Delta| > (m_{-} + m_{+})$ at zero displacement field, $E_{\perp} =0$.
% External potential breaks the mirror symmetry as shown in Fig 1c (dashed line).
%and 
%\begin{equation}
%\mathcal{H}^{(1)}_{+} = 
%\frac{\hbar^2 v^2}{2 \gamma^2_{1}}
%\bigg( A (\pi^{\dagger} \pi + \pi \pi^{\dagger}) + B \hat{\sigma}_{z} (\pi^{\dagger} \pi - \pi \pi^{\dagger})  \bigg),
%\bigg( (\Gamma_{1}\pi^{\dagger}\pi + \Gamma_{2} \pi \pi^{\dagger}) \mathcal{I} +  \hat{\sigma}_{z} (\Gamma_{1}\pi^{\dagger}\pi - \Gamma_{2} \pi \pi^{\dagger}) \bigg),
%\end{equation}
%where $\Gamma_{1}=13.5$ meV, %\delta + \gamma_{5}/4$,
 %$\Gamma_{2}=16$ meV (see Ref.~\onlinecite{PhysRevLett.117.076807}). %\gamma_{5}/4$ 
%To ensure the semi-metallic FS topology, plotted in Fig 1c, electron-hole band overlap at the CNP requires $|\Delta| > (m_{-} + m_{+})$. The band dispersion in the presence of an external potential, which breaks the mirror symmetry, is also plotted in Fig 1c (dashed line).

The LL spectrum  of ABA-TLG, plotted in Fig~\ref{figuretwo} b, shows that only the energies of the zeroth ($N=0$) LLs, are magnetic field independent~\cite{PhysRevB.87.115422,PhysRevLett.117.076807}. The odd and even parity zLLs, including the spin and valley degeneracy, have four and eight flavor components respectively. A common feature of $\mathcal{H}_{+}$ is the additional degeneracy associated with the $n=0,1$ LL orbitals in the even parity zLLs~\cite{PhysRevLett.96.086805}. These even-parity zLL orbitals are further split by $\Delta_{LL} \sim 0.24 B$ meV~\footnote{Coulomb interactions lead to $n = 0$ LL orbital polarization for a partially filled zeroth LL in bilayer graphene, see Y. Barlas {\it et. al.}, Phys. Rev. Lett. {\bf 101}, 097601 (2008).}. Additionally, lack of inversion symmetry in ABA-TLG breaks the valley degeneracy. The valley splitting is determined by the mass terms $m_{+}(m_{-})$, for the even and odd parity zLLs, respectively. Due to the energy gap, $\tilde{\Delta} = \Delta -(m_+ + m_-) >0$, the odd-parity zLLs (denoted by red) lie above the even-parity zLLs (denoted by blue).
% This results in counter-propagating edges states at the CNP associated with well defined parity.
 
\begin{figure}
\begin{tabular}{cc}
\includegraphics[clip,width=0.85\linewidth]{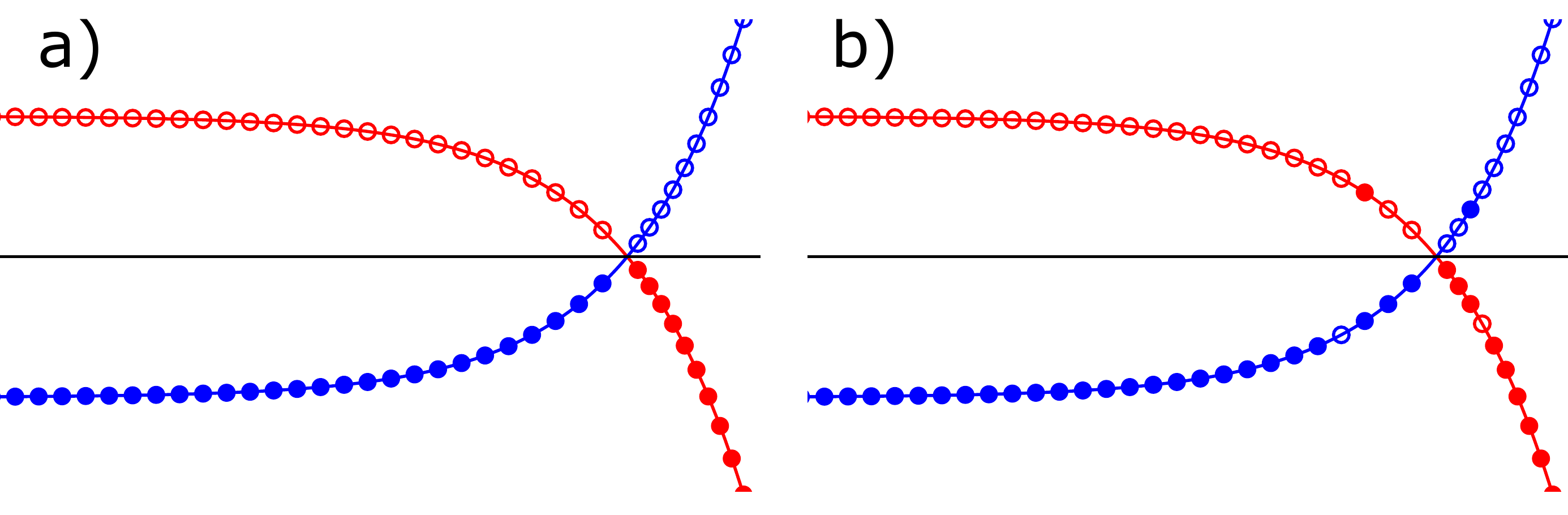}\\
\includegraphics[clip,width=1\linewidth]{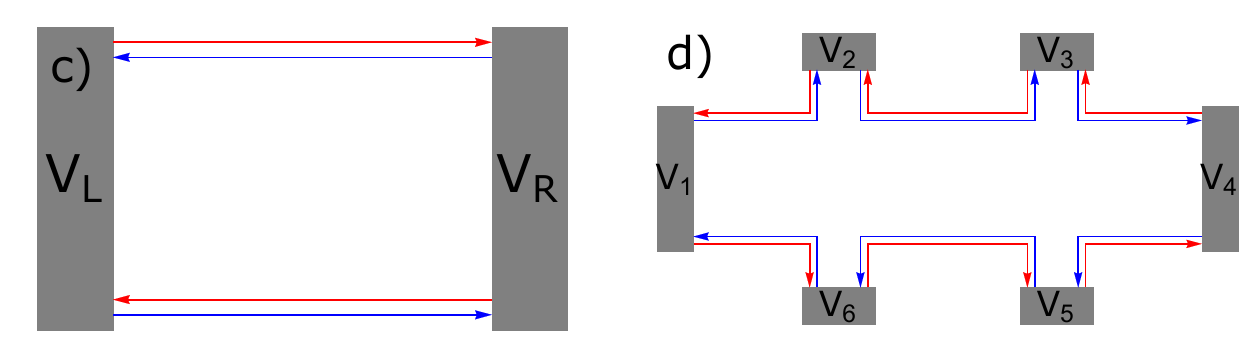}
\end{tabular}
\caption{a) and b) Edge dispersion of the $\nu = 1 + (-1)$ fractional parity Hall state, the counter-propagating edge currents flows due to particle-hole fluctuations in the even (blue) and odd (red) parity zLLs. Counter-propagating edge states of the fractional parity Hall effect in the two-terminal c) and Hall bar d) measurement setup.}
\label{figurethree} 
\end{figure}

The FS topology of ABA-TLG at the CNP (defined as $ \nu = \nu_{+} + \nu_{-} =0$) implies that above a very weak critical magnetic field $\nu_{-}=-2$ and $\nu_{+}= +2$ (including spin degeneracy), where $\nu_{\eta}$ denotes the filling factor of the $\eta$-parity zLLs). The edge states associated with the even-parity electron and odd-parity hole-zLLs flow in opposite directions, as depicted in Fig.~\ref{figurethree} (a) \& (b). The counter-propagating edges modes at $\nu=0$ are associated definite parity, which we call this the parity Hall effect.
One consequence of this state is that $\sigma_{xy}=0$ along with a quantized two-terminal longitudinal conductivity $\sigma_{xx} =4e^2/h$. Mirror symmetry prohibits back-scattering of the counter-propagating even and odd parity branches ensuring the quantization of $\sigma_{xx}$. %Consequently, the application of an electric field, which breaks the mirror symmetry, will lead to edge state localization %resulting in a rapid reduction of the longitudinal conductivity from $ 4e^2/h 
%$\sigma_{xx} \to 0$. 
This phase and its broken symmetry counterparts have been detected in dual-gate ABA-TLG samples~\footnote{C. N. Lau (private communication)}. %The rest of this paper focuses on the correlated states at the CNP with fractional filling factors $|\nu_{\pm}|=1/m $.

To proceed further, it is advantageous to express the zLL wavefunctions in the charge conjugation basis. Charge conjugation symmetry is defined as, 
\be
\label{chargeconjugation}
C_{\eta}^{\dagger} \mathcal{H}_{\eta}^{\star}(e) C_{\eta} = -\mathcal{H}_{\eta} (-e ),
\ee
where $C_{\eta} $ the charge conjugation symmetry operator is a unitary matrix, and $e$ is the electron charge.  It is easy to verify that for the odd parity Hamiltonian $\mathcal{H}_{-}$, the charge conjugation operator is: $C_{-} = \hat{\sigma}_{y} \hat{\sigma}_{z} $, while for the even parity Hamiltonian $\mathcal{H}_+$, the charge conjugation operator is: $ C_{+} =  \hat{\sigma}_{y}$. 
%Since the $C_{\eta}$ is block diagonal is the even and odd parity basis $[\hat{\mathcal{M}},C_{\eta} ] =0$, therefore the even and odd parity charges are good quantum numbers as long as mirror symmetry is preserved. This provides a natural basis for the zLL wavefunctions as we show below.   
So as long as the mirror symmetry is preserved, $[\mathcal{M},C_{\eta} ] =0$. One consequence of Eq.~\ref{chargeconjugation} is that, if $| \psi \rangle$ is an eigenstate of $\mathcal{H}_{\eta}$ with an eigenvalue $E_{\eta}$ and charge $e$, then $C_{\eta} | \psi \rangle^{\star}$ is also an eigenstate with the eigenvalue $-E_{\eta}$ and charge $-e$. As we see below, this charge conjugation symmetry provides a natural basis for the many-body wavefunctions in the parity Hall state.

In the symmetric gauge, ${\bm A}=B/2 (-y,x,0)$, the even parity $n=0$ orbital zLL electron-like and hole-like states in the charge conjugation basis become,
\be 
\label{evenparitystate}
\langle z| l,+,B_{2} \rangle = z^{l} e^{-\frac{1}{4} |z|^2}, \quad \langle z |l,+,A_{+}\rangle  = (z^{\star})^{l} e^{-\frac{1}{4} |z|^2},
\ee
where the electron-like states is localized on $B_{2}$ and the hole-like state is localized on $A_{+}$, $z=(x_+ +i y_+)/l_{B}$ denotes the position of the even parity electrons and $l$ is the angular momentum quantum number. The hole-like state in Eq.~\ref{evenparitystate} is calculated by applying $C_{+}$ to the conjugate of the even parity electron-like zLL wavefunction: $| l,+,B_{2} \rangle^{\star}$. Similarly, the odd-parity zLL hole-like and electron-like wavefunctions can be expressed as,  
\be
\langle w |l,-,A_{-}\rangle = (w^{\star})^{l} e^{-\frac{1}{4} |w|^2}, \quad \langle w |l,-,B_{-} \rangle = w^{l} e^{-\frac{1}{4} |w|^2},
\ee
which resides purely on the orbital $A_{-}$ and $B_{-}$ orbitals respectively, $w=(x_- +i y_-)/l_{B}$ denotes the position of the odd parity electrons. 
%As argued above, this prescription of labeling the even parity zLLs as electron and the odd parity zLLs as holes, only works in the presence of mirror symmetry.

The holomorphic and anti-holomorphic nature of the even and odd parity zLL lead to a Laughlin like class of correlated states~\cite{PhysRevLett.96.106802}, with fractional filling $|\nu_{\pm}|=1/m (m$ odd) at the CNP ($\nu = \nu_+ + \nu_- = 0$), 
\begin{equation}
\label{chargeconjugateLaughlin} 
\Psi_{0} = \prod_{i<j} (z_{i} - z_{j})^{m} (w^{\star}_{i} - w^{\star}_{j})^{m} \prod_{k} e^{-\frac{1}{4} (|z_{k}|^2 +|w_{k}|^2)}.  
\end{equation}
The correlated wavefunction above %is valid in the absence of interactions between the even and odd parity zLLs. 
satisfies mirror symmetry at $\nu = 0$, which requires invariance under $z_{i} \leftrightarrow w^{\star}_{i}$. The elementary excitations of $\Psi_{0}$, consists of a pair of quasi-electron/hole within each parity sector with fractional charge $e^{\star}_{\pm} = \nu_{\pm} e$~\cite{PhysRevLett.50.1395}. These excitations can be created by introducing an infinitesimal unit flux quanta at the position $x_{0}$~\cite{PhysRevLett.50.1395} with $\prod_{i} (z_{i} -x_{0}) (w^{\star}_{i} - x_{0}) \Psi_{0}$. 

$\Psi_{0}$ 
%minimizes the interaction energy by forming pairs with relative angular momentum $m$~\cite{PhysRevLett.51.605,PhysRevB.31.5280}. It
is the zero-energy eigenstate of the model Hamiltonian, $H = V_{l,\eta} P^{\eta, i}_{i,j} $ with $l = (m-1)/2$, where $V_{l,\eta} $ is the well-known Haldane pseudo-potential and $P^{\eta,l}_{i,j} $ is the projection operator on relative angular momentum $l$, for $\eta = \pm $ parity zLL~\cite{PhysRevLett.51.605,PhysRevB.31.5280}. 
This model Hamiltonian will coincide with the effective Hamiltonian at partial fillings as interactions, given by the energy scale $e^2/\epsilon l_{B} \sim (54/\epsilon) \sqrt{B} \  {\rm meV}$, renormalize the single-particle gap $\tilde{\Delta}$~\footnote{Interaction induced mixing with higher LLs will lead to positive/negative self-energy corrections to the electron-like/hole-like LLs resulting in a reduction of the single-particle gap $\tilde{\Delta}$}. In this case, the
%and an incompressible maximal density state at $|\nu_{\eta}|=1/m$ within each zLL. 
interaction induced gap, $\Delta_{e-e} $ can be estimated from previous studies of the fractional quantum Hall effect in graphene, for the $\nu=1/3$ state $\Delta_{e-e}\sim 0.035e^2/\epsilon l_{B} = (1.9/\epsilon) \sqrt{B} \  {\rm meV}$~\cite{PhysRevLett.113.086401}. The phase diagram for the fractional parity Hall effect and the hierarchy of counter-propagating fractional states will be discussed elsewhere~\footnote{Y. Barlas (in preparation)}.

%The correlated wavefunction above %is valid in the absence of interactions between the even and odd parity zLLs. 
%satisfies mirror symmetry at $\nu = 0$, which requires invariance under $z_{i} \leftrightarrow w^{\star}_{i}$. The elementary excitations of $\Psi_{0}$, consists of a pair of quasi-electron/hole within each parity sector with fractional charge $e^{\star}_{\pm} = \nu_{\pm} e$~\cite{PhysRevLett.50.1395}. These excitations can be created by introducing an infinitesimal unit flux quanta at the position $x_{0}$~\cite{PhysRevLett.50.1395} with $\prod_{i} (z_{i} -x_{0}) (w^{\star}_{i} - x_{0}) \Psi_{0}$. 

%Before addressing the transport properties, let's take a brief detour to look at the topological properties of the charge conjugate Laughlin state $\Psi_{0}$.
% {\bf (Explain the Lagrangian)}
The topological properties and excitations can be captured by the effective theory of the fractional parity Hall effect. It 
can be derived by performing the duality transformation~\footnote{see Supplemental Section [url] for the derivation, which includes 
Refs.~\onlinecite{PhysRevLett.58.1252,PhysRevLett.62.86,PhysRevLett.62.82,PhysRevLett.66.1220,PhysRevB.41.9377}} on the Chern-Simons Landau Ginzburg (CSLG) action~\cite{PhysRevLett.58.1252,PhysRevLett.62.86,PhysRevLett.62.82} describing the charge conjugate Laughlin state $\Psi_0$
(see supplemental section). The field theory is described by the effective Lagrangian, 
\begin{equation}
\label{dualaction}
%\mathcal{L} = \sum_{\eta=\pm} \bigg( -\eta e \epsilon^{\mu \nu \rho} A_{\mu} \partial_{\nu} b_{\rho}^{\eta} - \eta e \theta_{\eta} \phi_{0} \epsilon^{\mu \nu \rho} b^{\eta}_{\mu} \partial_{\nu} b_{\rho}^{\eta} - J_{\mu}^{\eta} j_{\mu}^{\eta}+ \cdots \bigg),
\mathcal{L} =  -\eta e \epsilon^{\mu \nu \rho} A_{\mu} \partial_{\nu} b_{\rho}^{\eta} - \eta e \theta_{\eta} \phi_{0} \epsilon^{\mu \nu \rho} b^{\eta}_{\mu} \partial_{\nu} b_{\rho}^{\eta} - J_{\mu}^{\eta} j_{\mu}^{\eta}+ \cdots ,
% \\
%&-& 2\pi b_{\mu}^{\eta} j_{\mu}^{\eta}+ \cdots \bigg),
\end{equation}
where $\nu=0,1,2$, $\eta=\pm$ and Einstein's summation convention is implied. The gauge fields $b_{\mu}$ and $ A_{\mu}$ denote to the Chern Simons (CS) statistical and electromagnetic gauge fields, $\theta_{\eta}= \eta m \pi$ is the parameter angle related to the individual filling factor $|\nu_{\pm}|$ and $\phi_{0} = hc/e$ is the flux quantum. The even and odd quasiparticle currents, given by $j^{\mu}_{\eta}$, couple to the CS gauge field current $J^{\eta}_{\nu} = 2 \pi b^{\eta}_{\mu}$, carry fractional charge $e^{\star}_{\pm} = \nu_{\pm } e$ and obey fractional statistics. The ground state degeneracy of $\mathcal{L}$ on a closed surface of genus $g$ is $m^{g} \times m^{g}$, due to the contribution from each parity sector~\cite{PhysRevB.41.9377}. In Eq.~\ref{dualaction}, mirror symmetry at $\nu =0$, given by $b_{\eta} \to b_{-\eta}$ and $\eta \to -\eta$, implies that $\sigma_{xy} =0$. However, as we show below the fractional parity Hall state exhibits a fractionally quantized two-terminal longitudinal conductance.

Now, we turn to the question of the precise longitudinal quantization of the charge conjugate Laughlin state. For the non-interacting $m=1$ charge conjugate Laughlin state, the edge excitations can be described in terms of single particle-hole excitations of the edge modes (see Fig.~\ref{figurethree}. The edge excitations consist of one-dimensional counter-propagating even and odd parity electron modes. In the presence of interactions, and for $m>1$ this single-particle description fails and the edge states must be defined as one dimensional bosonic density excitations~\cite{PhysRevB.41.12838} along the edge. In terms of the bosonic field $\phi_{\eta}$, the imaginary time action of the edge modes for $\Psi_{0}$ is given by,
\be
S_{0} = \frac{1}{4 \pi |\nu_{\pm}|} \sum_{\eta =e,o}\int dx d \tau \partial_{x} \phi_{\eta} (i \eta \partial_{\tau} + v \partial_{x}) \phi_{\eta},
\ee 
where the edge charge density is given by $\rho_{\eta}(x) = 1/(2 \pi) \partial_{x} \phi_{\eta}$. The first terms encodes the Kac-Moody commutation relations $[\phi_{\eta}(x), \phi_{\eta'} (x')] = i \pi \nu_{\eta} sgn(x-x') \delta_{\eta,\eta'}$.
We assume the velocity parameter $v$, determined by the intra-edge interactions and edge potential, is the same for both edge modes. Short-ranged inter-edge interactions, which can be expressed as,
\be
S_{int} = \frac{v_{d}}{2 \pi |\nu_{\pm}|} \int dx d \tau \partial_{x} \phi_{e} \partial_{x}\phi_{o},
\ee
result in equilibration of the edge modes. Their effect can be captured by diagonalizing the action $ S =S_{0} + S_{int}$.
This is done by introducing chiral bosonic fields, $\varphi_{L(R)} = a_{\mp} \phi_{e} + a_{\pm} \phi_{o}$ with $a_{\pm} = (\gamma \pm 1)/(2\sqrt{\gamma})$ where $\gamma =\sqrt{v_{+}/v_{-}}$ and $v_{\pm} = v \pm v_{d}$. In terms of the chiral fields $\varphi_{L(R)}$ the bosonized imaginary time action becomes,
\be 
S = \frac{1}{4 \pi |\nu_{\pm}|} \sum_{\alpha = L(R)} \int dx d \tau \big[ \partial_{x} \varphi_{\alpha} (i sgn(\alpha) \partial_{\tau} + u \partial_{x}) \varphi_{\alpha} \big] %+ \partial_{x} \varphi_{R} (i \partial_{\tau} + u \partial_{x}) \varphi_{R} \big],
\ee 
where $u =\sqrt{v_{+} v_{-}}=\sqrt{v^2-v_d^2}$ and $sgn(\alpha)= \pm$ for the $R(L)$ chiral fields. In terms of the chiral fields the edge modes of $\Psi_{0}$ are just two decoupled chiral Luttinger liquids~\cite{fisherkane1997}.

The one dimensional edge current is obtained from continuity equation, $I_{L(R)} = \dot{\varphi}_{L(R)}/(2\pi)$. When coupled to the electric potential $S \to S + \int dx d\tau \rho(x) V$, the transport characteristics of $S$ can be calculated in linear response~\cite{fisherkane1997}. Consider the chiral edge modes flowing between two-reservoirs at different chemical potentials, as indicated in Fig.~\ref{figurethree} (c). Using the Kubo formula for conductivity the current flowing along the top is given by, $I_{t} = g(V_{1}-V_{2})$ and at the bottom $I_{b} = g(V_{2}-V_{1})$ with $g = |\nu_{\pm}| e^2/h$. This gives the net current,  
\be
\label{2tertransport}
I=I_{t} - I_{b} = 2 |\nu_{\pm}| \frac{e^2}{h} (V_{L} -V_{R}),
\ee
resulting in two-terminal conductance $G = 2 |\nu_{\pm}| e^2/h$, independent of the velocity. It is then straightforward to generalize the above results to calculate the resistance in a Hall bar geometry, see Fig.~\ref{figurethree} (d). Using the multi-terminal Landuer-Buttiker theory for a Hall bar geometry, with a voltage applied between leads 1 and 4, we find zero Hall resistance, $R_{14,26} =0$, and a fractionally quantized longitudinal resistance, $R_{14,14} = h/(2 |\nu_{\pm}| e^2)$.

The effect of random edge disorder can be characterized by a localization length. This localization length has a power-law dependence on the strength of the disorder~\cite{fisherkane1997}. It determines the sample dimensions for which the quantized longitudinal conductivity can be observed. If the distance between the leads in Fig.~\ref{figurethree} is smaller than the localization length, the system will be equivalent to a disorder-free system and the conductance will be given by (\ref{2tertransport}). For rough edges, sharp local gates can be used to move the physical edge inside the sample thus preserving the requirement of mirror symmetry along the sample edges. 

%{Talk about the hierarchial states at other fractional filling factors $\nu$ and can be constructed from the well known schemes, and also dicuss the possibility of coexistence of the fractional Hall and fractional longitudinal conductance}.
The fractional parity Hall states proposed here can be detected in Hall measurements. With the chemical potential pinned to the charge neutrality point, the individual filling factors $\nu_{\pm} = 2 \pi n_{\pm} l^2_{B} \propto 1/B$ can be modified by increasing the magnetic field. As the field is increased, plateaus in the longitudinal conductivity at $2 |\nu_{\pm}| e^2/h$ should appear at certain fractional fillings $|\nu_{\pm}|=1/m$ in each parity sector, for sufficiently clean samples. Since the parity Hall states require mirror symmetry they are gate tunable.
 %Additionally, when the filling factors $\nu $ lies between $(2,-2)$, quantum Hall states with simultaneous quantized Hall and longitudinal resistances can be realized at $\nu=(m_2-m_1)/(m_1 m_2)$, ($m_1$ and $m_2$ odd). This should give $\sigma_{xy} = \nu e^2/h$ and $\sigma_{xx} = 2 \nu e^2/h$. 
At other partial fillings, exchange interactions will lead to a large class of magnetically ordered states. Such a state will exhibit spin-polarized counter-propagating edge modes with a gate tunable fractionally quantized longitudinal resistance. 
%This opens the possibility of using the edge states of ABA-TLG at $\nu =0$ to realize non-Abelian anyons~\cite{NatureComm.4.1348}.  

{\em Acknowledgements:}
The author would like to acknowledge discussions with C. N. Lau and Allan H. MacDonald 
along with E. Prodan, E. A. Henriksen and K. Yang for comments and suggestions on the manuscript. 
This work was partially supported by SHINES, an Energy Frontier Research Center, funded by the U.S. Department of Energy,
Office of Science, Basic Energy Sciences under Award \#DE-SC0012670.

%\bibliography{FMHbiblio}

\begin{thebibliography}{32}%
\makeatletter
\providecommand \@ifxundefined [1]{%
 \@ifx{#1\undefined}
}%
\providecommand \@ifnum [1]{%
 \ifnum #1\expandafter \@firstoftwo
 \else \expandafter \@secondoftwo
 \fi
}%
\providecommand \@ifx [1]{%
 \ifx #1\expandafter \@firstoftwo
 \else \expandafter \@secondoftwo
 \fi
}%
\providecommand \natexlab [1]{#1}%
\providecommand \enquote  [1]{``#1''}%
\providecommand \bibnamefont  [1]{#1}%
\providecommand \bibfnamefont [1]{#1}%
\providecommand \citenamefont [1]{#1}%
\providecommand \href@noop [0]{\@secondoftwo}%
\providecommand \href [0]{\begingroup \@sanitize@url \@href}%
\providecommand \@href[1]{\@@startlink{#1}\@@href}%
\providecommand \@@href[1]{\endgroup#1\@@endlink}%
\providecommand \@sanitize@url [0]{\catcode `\\12\catcode `\$12\catcode
  `\&12\catcode `\#12\catcode `\^12\catcode `\_12\catcode `\%12\relax}%
\providecommand \@@startlink[1]{}%
\providecommand \@@endlink[0]{}%
\providecommand \url  [0]{\begingroup\@sanitize@url \@url }%
\providecommand \@url [1]{\endgroup\@href {#1}{\urlprefix }}%
\providecommand \urlprefix  [0]{URL }%
\providecommand \Eprint [0]{\href }%
\providecommand \doibase [0]{http://dx.doi.org/}%
\providecommand \selectlanguage [0]{\@gobble}%
\providecommand \bibinfo  [0]{\@secondoftwo}%
\providecommand \bibfield  [0]{\@secondoftwo}%
\providecommand \translation [1]{[#1]}%
\providecommand \BibitemOpen [0]{}%
\providecommand \bibitemStop [0]{}%
\providecommand \bibitemNoStop [0]{.\EOS\space}%
\providecommand \EOS [0]{\spacefactor3000\relax}%
\providecommand \BibitemShut  [1]{\csname bibitem#1\endcsname}%
\let\auto@bib@innerbib\@empty
%</preamble>
\bibitem [{\citenamefont {Kane}\ and\ \citenamefont
  {Mele}(2005)}]{PhysRevLett.95.226801}%
  \BibitemOpen
  \bibfield  {author} {\bibinfo {author} {\bibfnamefont {C.~L.}\ \bibnamefont
  {Kane}}\ and\ \bibinfo {author} {\bibfnamefont {E.~J.}\ \bibnamefont
  {Mele}},\ }\href {\doibase 10.1103/PhysRevLett.95.226801} {\bibfield
  {journal} {\bibinfo  {journal} {Phys. Rev. Lett.}\ }\textbf {\bibinfo
  {volume} {95}},\ \bibinfo {pages} {226801} (\bibinfo {year}
  {2005})}\BibitemShut {NoStop}%
\bibitem [{\citenamefont {Bernevig}\ and\ \citenamefont
  {Zhang}(2006)}]{PhysRevLett.96.106802}%
  \BibitemOpen
  \bibfield  {author} {\bibinfo {author} {\bibfnamefont {B.~A.}\ \bibnamefont
  {Bernevig}}\ and\ \bibinfo {author} {\bibfnamefont {S.-C.}\ \bibnamefont
  {Zhang}},\ }\href {\doibase 10.1103/PhysRevLett.96.106802} {\bibfield
  {journal} {\bibinfo  {journal} {Phys. Rev. Lett.}\ }\textbf {\bibinfo
  {volume} {96}},\ \bibinfo {pages} {106802} (\bibinfo {year}
  {2006})}\BibitemShut {NoStop}%
\bibitem [{\citenamefont {K{\"o}nig}\ \emph {et~al.}(2007)\citenamefont
  {K{\"o}nig}, \citenamefont {Wiedmann}, \citenamefont {Br{\"u}ne},
  \citenamefont {Roth}, \citenamefont {Buhmann}, \citenamefont {Molenkamp},
  \citenamefont {Qi},\ and\ \citenamefont {Zhang}}]{Konig766}%
  \BibitemOpen
  \bibfield  {author} {\bibinfo {author} {\bibfnamefont {M.}~\bibnamefont
  {K{\"o}nig}}, \bibinfo {author} {\bibfnamefont {S.}~\bibnamefont {Wiedmann}},
  \bibinfo {author} {\bibfnamefont {C.}~\bibnamefont {Br{\"u}ne}}, \bibinfo
  {author} {\bibfnamefont {A.}~\bibnamefont {Roth}}, \bibinfo {author}
  {\bibfnamefont {H.}~\bibnamefont {Buhmann}}, \bibinfo {author} {\bibfnamefont
  {L.~W.}\ \bibnamefont {Molenkamp}}, \bibinfo {author} {\bibfnamefont {X.-L.}\
  \bibnamefont {Qi}}, \ and\ \bibinfo {author} {\bibfnamefont {S.-C.}\
  \bibnamefont {Zhang}},\ }\href {\doibase 10.1126/science.1148047} {\bibfield
  {journal} {\bibinfo  {journal} {Science}\ }\textbf {\bibinfo {volume}
  {318}},\ \bibinfo {pages} {766} (\bibinfo {year} {2007})}\BibitemShut
  {NoStop}%
\bibitem [{\citenamefont {Kharitonov}(2012)}]{PhysRevB.85.155439}%
  \BibitemOpen
  \bibfield  {author} {\bibinfo {author} {\bibfnamefont {M.}~\bibnamefont
  {Kharitonov}},\ }\href {\doibase 10.1103/PhysRevB.85.155439} {\bibfield
  {journal} {\bibinfo  {journal} {Phys. Rev. B}\ }\textbf {\bibinfo {volume}
  {85}},\ \bibinfo {pages} {155439} (\bibinfo {year} {2012})}\BibitemShut
  {NoStop}%
\bibitem [{\citenamefont {Young}\ \emph {et~al.}(2014)\citenamefont {Young},
  \citenamefont {Sanchez-Yamagishi}, \citenamefont {Hunt}, \citenamefont
  {Choi}, \citenamefont {Watanabe}, \citenamefont {Taniguchi}, \citenamefont
  {Ashoori},\ and\ \citenamefont {Jarillo-Herrero}}]{Young2014}%
  \BibitemOpen
  \bibfield  {author} {\bibinfo {author} {\bibfnamefont {A.~F.}\ \bibnamefont
  {Young}}, \bibinfo {author} {\bibfnamefont {J.~D.}\ \bibnamefont
  {Sanchez-Yamagishi}}, \bibinfo {author} {\bibfnamefont {B.}~\bibnamefont
  {Hunt}}, \bibinfo {author} {\bibfnamefont {S.~H.}\ \bibnamefont {Choi}},
  \bibinfo {author} {\bibfnamefont {K.}~\bibnamefont {Watanabe}}, \bibinfo
  {author} {\bibfnamefont {T.}~\bibnamefont {Taniguchi}}, \bibinfo {author}
  {\bibfnamefont {R.~C.}\ \bibnamefont {Ashoori}}, \ and\ \bibinfo {author}
  {\bibfnamefont {P.}~\bibnamefont {Jarillo-Herrero}},\ }\href
  {http://dx.doi.org/10.1038/nature12800} {\bibfield  {journal} {\bibinfo
  {journal} {Nature}\ }\textbf {\bibinfo {volume} {505}},\ \bibinfo {pages}
  {528} (\bibinfo {year} {2014})}\BibitemShut {NoStop}%
\bibitem [{\citenamefont {Maher}\ \emph {et~al.}(2013)\citenamefont {Maher},
  \citenamefont {Dean}, \citenamefont {Young}, \citenamefont {Taniguchi},
  \citenamefont {Watanabe}, \citenamefont {Shepard}, \citenamefont {Hone},\
  and\ \citenamefont {Kim}}]{Maher2013}%
  \BibitemOpen
  \bibfield  {author} {\bibinfo {author} {\bibfnamefont {P.}~\bibnamefont
  {Maher}}, \bibinfo {author} {\bibfnamefont {C.~R.}\ \bibnamefont {Dean}},
  \bibinfo {author} {\bibfnamefont {A.~F.}\ \bibnamefont {Young}}, \bibinfo
  {author} {\bibfnamefont {T.}~\bibnamefont {Taniguchi}}, \bibinfo {author}
  {\bibfnamefont {K.}~\bibnamefont {Watanabe}}, \bibinfo {author}
  {\bibfnamefont {K.~L.}\ \bibnamefont {Shepard}}, \bibinfo {author}
  {\bibfnamefont {J.}~\bibnamefont {Hone}}, \ and\ \bibinfo {author}
  {\bibfnamefont {P.}~\bibnamefont {Kim}},\ }\href {\doibase 10.1038/nphys2528}
  {\bibfield  {journal} {\bibinfo  {journal} {Nat Phys}\ }\textbf {\bibinfo
  {volume} {9}},\ \bibinfo {pages} {154} (\bibinfo {year} {2013})}\BibitemShut
  {NoStop}%
\bibitem [{\citenamefont {Levin}\ and\ \citenamefont
  {Stern}(2009)}]{PhysRevLett.103.196803}%
  \BibitemOpen
  \bibfield  {author} {\bibinfo {author} {\bibfnamefont {M.}~\bibnamefont
  {Levin}}\ and\ \bibinfo {author} {\bibfnamefont {A.}~\bibnamefont {Stern}},\
  }\href {\doibase 10.1103/PhysRevLett.103.196803} {\bibfield  {journal}
  {\bibinfo  {journal} {Phys. Rev. Lett.}\ }\textbf {\bibinfo {volume} {103}},\
  \bibinfo {pages} {196803} (\bibinfo {year} {2009})}\BibitemShut {NoStop}%
\bibitem [{\citenamefont {Henriksen}\ \emph {et~al.}(2012)\citenamefont
  {Henriksen}, \citenamefont {Nandi},\ and\ \citenamefont
  {Eisenstein}}]{PhysRevX.2.011004}%
  \BibitemOpen
  \bibfield  {author} {\bibinfo {author} {\bibfnamefont {E.~A.}\ \bibnamefont
  {Henriksen}}, \bibinfo {author} {\bibfnamefont {D.}~\bibnamefont {Nandi}}, \
  and\ \bibinfo {author} {\bibfnamefont {J.~P.}\ \bibnamefont {Eisenstein}},\
  }\href {\doibase 10.1103/PhysRevX.2.011004} {\bibfield  {journal} {\bibinfo
  {journal} {Phys. Rev. X}\ }\textbf {\bibinfo {volume} {2}},\ \bibinfo {pages}
  {011004} (\bibinfo {year} {2012})}\BibitemShut {NoStop}%
\bibitem [{\citenamefont {Koshino}\ and\ \citenamefont
  {McCann}(2010)}]{PhysRevB.81.115315}%
  \BibitemOpen
  \bibfield  {author} {\bibinfo {author} {\bibfnamefont {M.}~\bibnamefont
  {Koshino}}\ and\ \bibinfo {author} {\bibfnamefont {E.}~\bibnamefont
  {McCann}},\ }\href {\doibase 10.1103/PhysRevB.81.115315} {\bibfield
  {journal} {\bibinfo  {journal} {Phys. Rev. B}\ }\textbf {\bibinfo {volume}
  {81}},\ \bibinfo {pages} {115315} (\bibinfo {year} {2010})}\BibitemShut
  {NoStop}%
\bibitem [{\citenamefont {Latil}\ and\ \citenamefont
  {Henrard}(2006)}]{PhysRevLett.97.036803}%
  \BibitemOpen
  \bibfield  {author} {\bibinfo {author} {\bibfnamefont {S.}~\bibnamefont
  {Latil}}\ and\ \bibinfo {author} {\bibfnamefont {L.}~\bibnamefont
  {Henrard}},\ }\href {\doibase 10.1103/PhysRevLett.97.036803} {\bibfield
  {journal} {\bibinfo  {journal} {Phys. Rev. Lett.}\ }\textbf {\bibinfo
  {volume} {97}},\ \bibinfo {pages} {036803} (\bibinfo {year}
  {2006})}\BibitemShut {NoStop}%
\bibitem [{\citenamefont {Guinea}\ \emph {et~al.}(2006)\citenamefont {Guinea},
  \citenamefont {Castro~Neto},\ and\ \citenamefont
  {Peres}}]{PhysRevB.73.245426}%
  \BibitemOpen
  \bibfield  {author} {\bibinfo {author} {\bibfnamefont {F.}~\bibnamefont
  {Guinea}}, \bibinfo {author} {\bibfnamefont {A.~H.}\ \bibnamefont
  {Castro~Neto}}, \ and\ \bibinfo {author} {\bibfnamefont {N.~M.~R.}\
  \bibnamefont {Peres}},\ }\href {\doibase 10.1103/PhysRevB.73.245426}
  {\bibfield  {journal} {\bibinfo  {journal} {Phys. Rev. B}\ }\textbf {\bibinfo
  {volume} {73}},\ \bibinfo {pages} {245426} (\bibinfo {year}
  {2006})}\BibitemShut {NoStop}%
\bibitem [{\citenamefont {Min}\ and\ \citenamefont
  {MacDonald}(2008)}]{PhysRevB.77.155416}%
  \BibitemOpen
  \bibfield  {author} {\bibinfo {author} {\bibfnamefont {H.}~\bibnamefont
  {Min}}\ and\ \bibinfo {author} {\bibfnamefont {A.~H.}\ \bibnamefont
  {MacDonald}},\ }\href {\doibase 10.1103/PhysRevB.77.155416} {\bibfield
  {journal} {\bibinfo  {journal} {Phys. Rev. B}\ }\textbf {\bibinfo {volume}
  {77}},\ \bibinfo {pages} {155416} (\bibinfo {year} {2008})}\BibitemShut
  {NoStop}%
\bibitem [{\citenamefont {Serbyn}\ and\ \citenamefont
  {Abanin}(2013)}]{PhysRevB.87.115422}%
  \BibitemOpen
  \bibfield  {author} {\bibinfo {author} {\bibfnamefont {M.}~\bibnamefont
  {Serbyn}}\ and\ \bibinfo {author} {\bibfnamefont {D.~A.}\ \bibnamefont
  {Abanin}},\ }\href {\doibase 10.1103/PhysRevB.87.115422} {\bibfield
  {journal} {\bibinfo  {journal} {Phys. Rev. B}\ }\textbf {\bibinfo {volume}
  {87}},\ \bibinfo {pages} {115422} (\bibinfo {year} {2013})}\BibitemShut
  {NoStop}%
\bibitem [{\citenamefont {Stepanov}\ \emph {et~al.}(2016)\citenamefont
  {Stepanov}, \citenamefont {Barlas}, \citenamefont {Espiritu}, \citenamefont
  {Che}, \citenamefont {Watanabe}, \citenamefont {Taniguchi}, \citenamefont
  {Smirnov},\ and\ \citenamefont {Lau}}]{PhysRevLett.117.076807}%
  \BibitemOpen
  \bibfield  {author} {\bibinfo {author} {\bibfnamefont {P.}~\bibnamefont
  {Stepanov}}, \bibinfo {author} {\bibfnamefont {Y.}~\bibnamefont {Barlas}},
  \bibinfo {author} {\bibfnamefont {T.}~\bibnamefont {Espiritu}}, \bibinfo
  {author} {\bibfnamefont {S.}~\bibnamefont {Che}}, \bibinfo {author}
  {\bibfnamefont {K.}~\bibnamefont {Watanabe}}, \bibinfo {author}
  {\bibfnamefont {T.}~\bibnamefont {Taniguchi}}, \bibinfo {author}
  {\bibfnamefont {D.}~\bibnamefont {Smirnov}}, \ and\ \bibinfo {author}
  {\bibfnamefont {C.~N.}\ \bibnamefont {Lau}},\ }\href {\doibase
  10.1103/PhysRevLett.117.076807} {\bibfield  {journal} {\bibinfo  {journal}
  {Phys. Rev. Lett.}\ }\textbf {\bibinfo {volume} {117}},\ \bibinfo {pages}
  {076807} (\bibinfo {year} {2016})}\BibitemShut {NoStop}%
\bibitem [{Note1()}]{Note1}%
  \BibitemOpen
  \bibinfo {note} {To keep the discussion simple, we express $\protect \mathcal
  {H}_{+}$ within second order perturbation theory.}\BibitemShut {Stop}%
\bibitem [{\citenamefont {McCann}\ and\ \citenamefont
  {Fal'ko}(2006)}]{PhysRevLett.96.086805}%
  \BibitemOpen
  \bibfield  {author} {\bibinfo {author} {\bibfnamefont {E.}~\bibnamefont
  {McCann}}\ and\ \bibinfo {author} {\bibfnamefont {V.~I.}\ \bibnamefont
  {Fal'ko}},\ }\href {\doibase 10.1103/PhysRevLett.96.086805} {\bibfield
  {journal} {\bibinfo  {journal} {Phys. Rev. Lett.}\ }\textbf {\bibinfo
  {volume} {96}},\ \bibinfo {pages} {086805} (\bibinfo {year}
  {2006})}\BibitemShut {NoStop}%
\bibitem [{Note2()}]{Note2}%
  \BibitemOpen
  \bibinfo {note} {Coulomb interactions lead to $n = 0$ LL orbital polarization
  for a partially filled zeroth LL in bilayer graphene, see Y. Barlas {\protect
  \it et. al.}, Phys. Rev. Lett. {\protect \bf 101}, 097601
  (2008).}\BibitemShut {Stop}%
\bibitem [{Note3()}]{Note3}%
  \BibitemOpen
  \bibinfo {note} {C. N. Lau (private communication)}\BibitemShut {NoStop}%
\bibitem [{\citenamefont {Laughlin}(1983)}]{PhysRevLett.50.1395}%
  \BibitemOpen
  \bibfield  {author} {\bibinfo {author} {\bibfnamefont {R.~B.}\ \bibnamefont
  {Laughlin}},\ }\href {\doibase 10.1103/PhysRevLett.50.1395} {\bibfield
  {journal} {\bibinfo  {journal} {Phys. Rev. Lett.}\ }\textbf {\bibinfo
  {volume} {50}},\ \bibinfo {pages} {1395} (\bibinfo {year}
  {1983})}\BibitemShut {NoStop}%
\bibitem [{\citenamefont {Haldane}(1983)}]{PhysRevLett.51.605}%
  \BibitemOpen
  \bibfield  {author} {\bibinfo {author} {\bibfnamefont {F.~D.~M.}\
  \bibnamefont {Haldane}},\ }\href {\doibase 10.1103/PhysRevLett.51.605}
  {\bibfield  {journal} {\bibinfo  {journal} {Phys. Rev. Lett.}\ }\textbf
  {\bibinfo {volume} {51}},\ \bibinfo {pages} {605} (\bibinfo {year}
  {1983})}\BibitemShut {NoStop}%
\bibitem [{\citenamefont {Trugman}\ and\ \citenamefont
  {Kivelson}(1985)}]{PhysRevB.31.5280}%
  \BibitemOpen
  \bibfield  {author} {\bibinfo {author} {\bibfnamefont {S.~A.}\ \bibnamefont
  {Trugman}}\ and\ \bibinfo {author} {\bibfnamefont {S.}~\bibnamefont
  {Kivelson}},\ }\href {\doibase 10.1103/PhysRevB.31.5280} {\bibfield
  {journal} {\bibinfo  {journal} {Phys. Rev. B}\ }\textbf {\bibinfo {volume}
  {31}},\ \bibinfo {pages} {5280} (\bibinfo {year} {1985})}\BibitemShut
  {NoStop}%
\bibitem [{Note4()}]{Note4}%
  \BibitemOpen
  \bibinfo {note} {Interaction induced mixing with higher LLs will lead to
  positive/negative self-energy corrections to the electron-like/hole-like LLs
  resulting in a reduction of the single-particle gap $\protect \mathaccentV
  {tilde}07E{\Delta }$}\BibitemShut {NoStop}%
\bibitem [{\citenamefont {Peterson}\ and\ \citenamefont
  {Nayak}(2014)}]{PhysRevLett.113.086401}%
  \BibitemOpen
  \bibfield  {author} {\bibinfo {author} {\bibfnamefont {M.~R.}\ \bibnamefont
  {Peterson}}\ and\ \bibinfo {author} {\bibfnamefont {C.}~\bibnamefont
  {Nayak}},\ }\href {\doibase 10.1103/PhysRevLett.113.086401} {\bibfield
  {journal} {\bibinfo  {journal} {Phys. Rev. Lett.}\ }\textbf {\bibinfo
  {volume} {113}},\ \bibinfo {pages} {086401} (\bibinfo {year}
  {2014})}\BibitemShut {NoStop}%
\bibitem [{Note5()}]{Note5}%
  \BibitemOpen
  \bibinfo {note} {Y. Barlas (in preparation)}\BibitemShut {NoStop}%
\bibitem [{Note6()}]{Note6}%
  \BibitemOpen
  \bibinfo {note} {See Supplemental Section [url] for the derivation, which
  includes Refs.~\protect \rev@citealp
  {PhysRevLett.58.1252,PhysRevLett.62.86,PhysRevLett.62.82,PhysRevLett.66.1220,PhysRevB.41.9377}}\BibitemShut
  {NoStop}%
\bibitem [{\citenamefont {Girvin}\ and\ \citenamefont
  {MacDonald}(1987)}]{PhysRevLett.58.1252}%
  \BibitemOpen
  \bibfield  {author} {\bibinfo {author} {\bibfnamefont {S.~M.}\ \bibnamefont
  {Girvin}}\ and\ \bibinfo {author} {\bibfnamefont {A.~H.}\ \bibnamefont
  {MacDonald}},\ }\href {\doibase 10.1103/PhysRevLett.58.1252} {\bibfield
  {journal} {\bibinfo  {journal} {Phys. Rev. Lett.}\ }\textbf {\bibinfo
  {volume} {58}},\ \bibinfo {pages} {1252} (\bibinfo {year}
  {1987})}\BibitemShut {NoStop}%
\bibitem [{\citenamefont {Read}(1989)}]{PhysRevLett.62.86}%
  \BibitemOpen
  \bibfield  {author} {\bibinfo {author} {\bibfnamefont {N.}~\bibnamefont
  {Read}},\ }\href {\doibase 10.1103/PhysRevLett.62.86} {\bibfield  {journal}
  {\bibinfo  {journal} {Phys. Rev. Lett.}\ }\textbf {\bibinfo {volume} {62}},\
  \bibinfo {pages} {86} (\bibinfo {year} {1989})}\BibitemShut {NoStop}%
\bibitem [{\citenamefont {Zhang}\ \emph {et~al.}(1989)\citenamefont {Zhang},
  \citenamefont {Hansson},\ and\ \citenamefont {Kivelson}}]{PhysRevLett.62.82}%
  \BibitemOpen
  \bibfield  {author} {\bibinfo {author} {\bibfnamefont {S.~C.}\ \bibnamefont
  {Zhang}}, \bibinfo {author} {\bibfnamefont {T.~H.}\ \bibnamefont {Hansson}},
  \ and\ \bibinfo {author} {\bibfnamefont {S.}~\bibnamefont {Kivelson}},\
  }\href {\doibase 10.1103/PhysRevLett.62.82} {\bibfield  {journal} {\bibinfo
  {journal} {Phys. Rev. Lett.}\ }\textbf {\bibinfo {volume} {62}},\ \bibinfo
  {pages} {82} (\bibinfo {year} {1989})}\BibitemShut {NoStop}%
\bibitem [{\citenamefont {Wen}\ and\ \citenamefont
  {Niu}(1990)}]{PhysRevB.41.9377}%
  \BibitemOpen
  \bibfield  {author} {\bibinfo {author} {\bibfnamefont {X.~G.}\ \bibnamefont
  {Wen}}\ and\ \bibinfo {author} {\bibfnamefont {Q.}~\bibnamefont {Niu}},\
  }\href {\doibase 10.1103/PhysRevB.41.9377} {\bibfield  {journal} {\bibinfo
  {journal} {Phys. Rev. B}\ }\textbf {\bibinfo {volume} {41}},\ \bibinfo
  {pages} {9377} (\bibinfo {year} {1990})}\BibitemShut {NoStop}%
\bibitem [{\citenamefont {Wen}(1990)}]{PhysRevB.41.12838}%
  \BibitemOpen
  \bibfield  {author} {\bibinfo {author} {\bibfnamefont {X.~G.}\ \bibnamefont
  {Wen}},\ }\href {\doibase 10.1103/PhysRevB.41.12838} {\bibfield  {journal}
  {\bibinfo  {journal} {Phys. Rev. B}\ }\textbf {\bibinfo {volume} {41}},\
  \bibinfo {pages} {12838} (\bibinfo {year} {1990})}\BibitemShut {NoStop}%
\bibitem [{\citenamefont {Kane}\ and\ \citenamefont
  {Fisher}(1997)}]{fisherkane1997}%
  \BibitemOpen
  \bibfield  {author} {\bibinfo {author} {\bibfnamefont {C.~L.}\ \bibnamefont
  {Kane}}\ and\ \bibinfo {author} {\bibfnamefont {M.~P.~A.}\ \bibnamefont
  {Fisher}},\ }in\ \href@noop {} {\emph {\bibinfo {booktitle} {Perspectives in
  Quantum Hall Effects}}},\ \bibinfo {editor} {edited by\ \bibinfo {editor}
  {\bibfnamefont {S.~D.}\ \bibnamefont {Sarma}}\ and\ \bibinfo {editor}
  {\bibfnamefont {A.}~\bibnamefont {Pinzuk}}}\ (\bibinfo  {publisher} {John
  Wiley and Sons},\ \bibinfo {address} {John Wiley and Sons},\ \bibinfo {year}
  {1997})\ Chap.~\bibinfo {chapter} {4}, pp.\ \bibinfo {pages}
  {109--157}\BibitemShut {NoStop}%
\bibitem [{\citenamefont {Lee}\ and\ \citenamefont
  {Zhang}(1991)}]{PhysRevLett.66.1220}%
  \BibitemOpen
  \bibfield  {author} {\bibinfo {author} {\bibfnamefont {D.-H.}\ \bibnamefont
  {Lee}}\ and\ \bibinfo {author} {\bibfnamefont {S.-C.}\ \bibnamefont
  {Zhang}},\ }\href {\doibase 10.1103/PhysRevLett.66.1220} {\bibfield
  {journal} {\bibinfo  {journal} {Phys. Rev. Lett.}\ }\textbf {\bibinfo
  {volume} {66}},\ \bibinfo {pages} {1220} (\bibinfo {year}
  {1991})}\BibitemShut {NoStop}%
\end{thebibliography}
%\bibliographystyle{plain}
%

%\begin{thebibliography}{99}
\end{document}